\documentclass[aps,epsf,psfig,fleqn]{revtex4}

\newcommand{\bed}{\[}
\newcommand{\eed}{\]}
\newcommand{\beq}{\begin{equation}}
\newcommand{\eeq}{\end{equation}}
\newcommand{\beqa}{\begin{eqnarray}}
\newcommand{\eeqa}{\end{eqnarray}}
\newcommand{\ket} [1] {\vert #1 \rangle}
\newcommand{\bra} [1] {\langle #1 \vert}

\newcommand{\gras}[1]{\bold{#1}}

\newcommand{\sysid}[1]{\mathsf{#1}}
\newcommand{\Tr}{\mathop{\mathrm{Tr}}}
\newcommand{\tr}{\mathop{\mathrm{tr}}}

\usepackage[dvips]{graphicx}
\usepackage[mathscr]{eucal}
\usepackage{amsmath}
\usepackage{amssymb}

\begin{document}
\bibliographystyle{prsty}

\title{Security of Quantum Key Distribution with Coherent States and Homodyne Detection}
\author{S. Iblisdir, G. Van Assche and N. J. Cerf}
\affiliation{Ecole Polytechnique, CP 165/59, Universit\'e Libre de Bruxelles, 1050 Brussels, Belgium}
\email{{siblisdi,gvanassc,ncerf}@ulb.ac.be}

\date{\today}

\begin{abstract}
We assess the security of a quantum key distribution protocol relying on the transmission of Gaussian-modulated coherent states and homodyne detection. This protocol is shown to be equivalent to a squeezed state protocol based on a CSS code construction, and is thus provably secure against any eavesdropping strategy. We also briefly show how this protocol can be generalized in order to improve the net key rate.
\end{abstract}

\maketitle

\section{Introduction}

Quantum Key Distribution (QKD) uses quantum mechanics to provide two parties (Alice and Bob) with a secret key, which they can later use to encrypt confidential information \cite{gisi02:qc}. Unlike classical key distribution, QKD relies, at least in principle, on no computational assumption, but only draws its validity from the laws of quantum mechanics. The resources needed for QKD always comprise a source of non-orthogonal quantum states on Alice's side, a quantum channel conveying these states to Bob, a measuring apparatus on Bob's side, and a (public) authenticated classical channel between Alice and Bob. QKD protocols generally consist in two (intertwined) parts. One part consists in probing the quantum channel to determine whether it is possible to securely transmit the key over it. The use of non-orthogonal quantum states allows to achieve this task. The other part consists in the explicit distillation of the secret key.

Most interest in QKD has been devoted to protocols involving (an approximation to) a single-photon source on Alice's side and a single-photon detector on Bob's side \cite{gisi02:qc,beve02}. However, protocols involving quantum continuous variables have lately been considered with an increasing interest \cite{hill00,gott01:crypto,cerf01:qdgk,silb02,silb02:db}. Of special importance are ``coherent-state'' protocols \cite{gros02:coherent,gros03}. The quantum source at Alice's side then randomly generates coherent states of a light mode with a Gaussian distribution, while Bob performs homodyne measurements. These protocols are very important  because they seem to allow for facilitated implementations and much higher secret-key generation rates than the protocols involving single-photon sources \cite{gros03}.

In this paper, we will constructively prove that secure coherent-state protocols relying on homodyne detection exist. A first security analysis of coherent-state protocols has been carried in \cite{gros02:coherent,gros03}, but only individual Gaussian eavesdropping strategies were considered. We here want to address a more general setting and allow a potential eavesdropper (Eve) to probe the quantum channel between Alice and Bob in any manner she pleases. We want to establish the security of coherent-state protocols against arbitrary collective attacks (thereby extending \cite{gott01}). The importance of our result lies in that it shows that no non-classical feature of light, such as squeezing, is necessary in continuous-variable quantum cryptography: coherent states, homodyne detection, and well-chosen communication procedures are sufficient for Alice and Bob to securely distill a secret key. 

\section{Squeezed-state protocols}

The basic ingredient that we shall use in the remaining is the argument used in \cite{shor00} to prove that the BB84 protocol is secure, against any eavesdropping strategy, when the procedures used for error correction and privacy amplification are derived from a CSS quantum error-correcting code \cite{cald96,stea96}. Let us start with a brief review of this argument. 

It is well known that quantum error-correcting codes provide a means to perform entanglement purification with one-way communication \cite{benn96_mseqec}. If two parties, Alice and Bob, share $N$ noisy entangled qubit pairs, their situation is fully equivalent to a situation where Alice would have prepared $N$ pairs, all in the Einstein-Podolsky-Rosen (EPR) state:
\beq\label{eq:phiplus}
\ket{\phi^+}=\frac{1}{\sqrt{2}}(\ket{00}+\ket{11}),
\eeq
and would have kept half of each pair for herself while sending all other halves to Bob through some noisy quantum channel. The effect of this channel on the state can be modeled as if the state either remains unaltered or undergoes either one of the three following ``errors'': bit-flip, $\phi^+ \to \psi^+$, or phase-flip, $\phi^+ \to \phi^-$ or both, $\phi^+ \to \psi^-$, where $\ket{\phi^-}=\frac{1}{\sqrt{2}}(\ket{00}-\ket{11})$ and $\ket{\psi^{\pm}}=\frac{1}{\sqrt{2}}(\ket{01} \pm \ket{10})$. In the latter situation, Alice and Bob could get pure EPR pairs upon Alice using a quantum error correcting code (QECC) to protect the halves sent to Bob from the noise effected by the channel. Equivalently, in the former situation, Alice and Bob can get $C N$ pairs in the state (\ref{eq:phiplus}) ($C \leq 1$) upon Alice and Bob measuring the syndromes (or error patterns) of some QECC, Alice communicating the values of her syndromes to Bob, and Bob performing error correction so as to align the values of his syndromes on those of Alice. $C$ is then the rate of the used quantum code. It is trivial to achieve secure QKD from entanglement purification because if Alice and Bob share a 2-qubit system in the state (\ref{eq:phiplus}), they certainly can extract a secret bit from it.

A (binary) CSS code is a $2^k$-dimensional subspace of the Hilbert space of $n$ qubits ($k \leq n$). Such a code belongs to the class of so-called stabilizer codes, i.e. they are defined as the eigenspace of a set of mutually commutating operators $\{\mathscr{O}_1,\ldots,\mathscr{O}_X\}$, the stabilizer generators. The essential feature of a CSS code is that all stabilizer generators are either of the form $X^{s_1} \otimes \ldots \otimes X^{s_n}$ or of the form $Z^{s_1} \otimes \ldots \otimes Z^{s_n}$, where $X\ket{i}=\ket{i \oplus 1}, Z\ket{i}=(-)^{i}\ket{i}$, and where $(s_1,\ldots,s_n) \in \{0,1\}^n$. Because of this feature, it is possible to prove that entanglement purification using a CSS code followed by key extraction is fully equivalent to a quantum cryptographic protocol with BB84 as a physical part and suitable error correction and privacy amplification as classical post-processing part \cite{shor00}. These procedures read as follows. Let the binary vectors $\mathscr{K}$ and $\mathscr{K}'$ denote respectively Alice's and Bob's raw key bits, and let $\mathscr{C}_2 \subset \mathscr{C}_1$ denote two embedded $n$-bit classical linear codes, with parity check matrices respectively $H_1$ and $H_2$ \cite{cover}. Alice announces the syndrome $H_1 \mathscr{K}= \xi^b$. Bob corrects $\mathscr{K}'$ to the nearest vector $\mathscr{K}''$ such that $H_1 \mathscr{K}''= \xi^b$ (error correction). With high probability, $\mathscr{K}''=\mathscr{K}$. The key is then reduced to $H_2 \mathscr{K}$ (privacy amplification).

Entanglement purification using CSS codes is (asymptotically) achievable as long as the bit-flip probability $e_b$ and the phase-flip probability $e_p$ satisfy
\beq\label{eq:CSSrate}
C \equiv 1-h(e_b)-h(e_p) > 0,
\eeq
where $h(x)=-\log_2 x^x (1-x)^{(1-x)}$ denotes the binary Shannon entropy \cite{shor00}.
Equivalently, the BB84 protocol will allow Alice and Bob to distill a secret key using the error correction and privacy amplification we have described if the error rates for two conjugate bases satisfy Eq.(\ref{eq:CSSrate}).

From QKD schemes based on entanglement purification of qubits, it is possible to derive a secure QKD scheme using squeezed states and homodyne detection, which is in spirit very close to the BB84 protocol \cite{gott01:crypto}. Let us present this scheme in a slightly modified form. Let $\hat{x}$ and $\hat{p}$ denote two conjugate quadratures of a single mode of the electromagnetic field 
($[\hat{x},\hat{p}]=i$). Alice creates (about) $4N$ quantum oscillators in a squeezed state. She draws a $4N$-bit string $b$ to decide for each of the $4N$ oscillator whether it will be prepared in an $x$-squeezed state or in a $p$-squeezed state. Also, for each oscillator, she draws a real value $x$ (or $p$) according to a probability distribution $P_{\textrm{pos}}(x)$ (or $P_{\textrm{mom}}(p)$), and sends Bob an $x$- (or $p$-)squeezed state centered on $(x,0)$ (or $(0,p)$). Bob receives the states and decides at random to measure them either in the $x$-basis or in the $p$-basis. By public discussion, Alice and Bob discard the oscillators for which Alice's choice of preparation and Bob's choice of measurement don't match. Alice and Bob should now have a list of (about) $2N$ correlated real values $(x_1,x'_1) \ldots (x_{2N},x'_{2N})$ from which they wish to extract bits. To do so, they proceed as follows. For each real value, $x$, Alice decomposes $x$ as
\beq
\label{eq:xdecomposed}
x=(S(x)+ \bar{S}(x))\sqrt{\pi}
\eeq
where $S(x) \in \mathbf{Z}$, and reveals $\bar{S}(x)=\textrm{frac}(x/\sqrt{\pi})$ (or $\bar{S}(p)=\textrm{frac}(p/\sqrt{\pi})$) to Bob. Alice's bit is the parity of $S(x)$ (resp. $S(p)$). Bob subtracts $\bar{S}(x) \sqrt{\pi}$ from his corresponding real value, $x'$, and adjusts his result $x'-\bar{S}(x) \sqrt{\pi}$ to the nearest integer multiple of $\sqrt{\pi}$. The key bit will be $0$ if this integer is even, and $1$ otherwise. At this point, Alice and Bob agree on a subset of size (about) $N$ of their key elements that they use for verification. A bit error (resp. a phase error) occurs when Alice sends an $x$-squeezed state (resp. a $p$-squeezed state), and Alice's bit and Bob's bit mismatch. If the estimates of the error rates $e_b$ and $e_p$ satisfy Eq.(\ref{eq:CSSrate}), Alice and Bob further proceed with error correction and privacy amplification as described above, and distill a secret key.

Owing to the manner the real axis is binned to associate bits to real numbers, the bit  error rate $e_b$ is bounded by the probability that, when Alice sends an $x$-squeezed state centered on the value $x_0$, $\ket{\textrm{sq}(x_0)}$, and Bob performs an $\hat{x}$ homodyne measurement, Bob gets an outcome whose value differs from $x_0$ by a value greater than $\sqrt{\pi}/2$. The phase error rate, $e_p$, can be bounded similarly.
Therefore, even in the absence of eavesdropping, $e_b$ and $e_p$ will be nonzero, due to finite squeezing. Quantifying squeezing with $10 \log_{10} \frac{1}{\tilde{\sigma}^2}$, where the states sent by Alice read $\ket{\textrm{sq}(x_0)} \sim \int dx \; e^{-(x-x_0)^2/2 \tilde{\sigma}^2} \ket{x}$ and $\ket{\textrm{sq}(p_0)} \sim \int dp \; e^{-(p-p_0)^2/2 \tilde{\sigma}^2} \ket{p}$, it was proven in \cite{gott01:crypto} that a minimum of 2.51 dB of squeezing is necessary for the protocol to work.

\section{Conversion to coherent-state protocols}

\subsection{Asymmetric squeezed-state protocols}

A first step in the conversion to a coherent-state protocol is to observe that three modifications can be brought to the above squeezed-state protocol without weakening its security. First, as shown in \cite{gott01:crypto}, the above protocol is equivalent to a protocol where Alice reveals $\bar{S}(x)=\textrm{frac}(x/\alpha \sqrt{\pi})$ when using the $x$- quadrature, and $\bar{S}(p)=\textrm{frac}(p\alpha/\sqrt{\pi})$ when using the $p$ quadrature, where $\alpha$ is some positive real parameter. Such an asymmetric protocol allows Alice to squeeze unequally  $x$ and $p$ quadratures. The squeezing should only be such that Eq.(\ref{eq:CSSrate}) is obeyed. In particular, Alice can use coherent states when encoding in the $x$ quadrature, if when encoding in the $p$ quadrature, she uses a state exhibiting a squeezing of at least 3.37 dB. Our second observation concerns the method used by Alice for encoding. When she chooses to encode in the $x$-quadrature, she draws the value of $x$ from $P_{\textrm{pos}}$ and prepares a coherent state centered on $(x,0)$. Similarly, when encoding with the conjugate quadrature, she prepares $p$-squeezed states centered on $(0,p)$. The decision to prepare states centered on $(x,0)$ or $(0,p)$ relies on an arbitrary convention between Alice and Bob for the axis for $x$ quadrature and the axis for $p$ quadrature. Instead of sending a state centered on $(x,0)$ (resp. $(0,p)$), Alice could as well send a state centered on $(x,p)$, when the key information is encoded in $x$ (resp. in $p$), and where the value $p$ (resp. $x$), drawn from some probability distribution $P'_{\textrm{pos}}(p)$ (resp. $P'_{\textrm{mom}}(x)$), may in principle be publicly disclosed to allow Bob to re-translate the state on the $x$ (or $p$) axis. Finally, we remark that the protocol is no less secure if Alice and Bob decide that the key is only encoded in the coherent states and never in the squeezed states. They can decide that about half of the time, Alice will send coherent states to transmit the key and to estimate $e_b$, while about half of the time, Alice will send squeezed states to estimate $e_p$. This fact holds for BB84 as well: one can decide that the key is only encoded in $Z$ eigenstates, and that $X$ eigenstates are only sent to determine the phase error rate. As long as $e_b$ and $e_p$ satisfy Eq.(\ref{eq:CSSrate}), the protocol will work safely. 

In summary, the following is a secure protocol.

$\# 1$ Alice prepares the state
$S=S^{\textrm{coh}} \otimes S^{\textrm{sq}}$, where 
\beq\label{eq:defscoh}
S^{\textrm{coh}}=S_{\textrm{key}}\otimes S_{\textrm{ck}}^{\textrm{b}}.
\eeq
$S_{\textrm{key}}=\gamma{(1)} \otimes \ldots \gamma{(N)}$ is a tensor product of $N$ coherent states, each drawn from a probability distribution $P_{\textrm{pos}}(x) P'_{\textrm{pos}}(p)$. Also, $S_{\textrm{ck}}^{\textrm{b}}=\gamma_{\textrm{c}}(1) \otimes \ldots \otimes \gamma_{\textrm{c}}(\mu)$
is a tensor product of coherent states, drawn from the same probability distribution $P_{\textrm{pos}}(x) P'_{\textrm{pos}}(p)$, and $S^{\textrm{sq}}=\sigma(1) \otimes \ldots \otimes \sigma(\nu)$ is a tensor product of $p$-squeezed states drawn from some probability distribution  $P'_{\textrm{mom}}(x) P_{\textrm{mom}}(p) $.  The probability distributions  $P_{\textrm{pos}}(x) P'_{\textrm{pos}}(p)$ and $P'_{\textrm{mom}}(x) P_{\textrm{mom}}(p) $ are such that
\beq
\int \; dx \; dp \; P_{\textrm{pos}}(x) P'_{\textrm{pos}}(p) \; \gamma(x,p)=
\int \; dx \; dp \; P'_{\textrm{mom}}(x) P_{\textrm{mom}}(p)  \; \sigma(x,p),
\eeq 
where $\gamma(x,p)$ (resp. $\sigma(x,p)$) denotes a coherent (resp. $p$-squeezed) state centered on $(x,p)$. 

$\# 2$ Alice picks a random permutation $\pi \in \textrm{Sym}(g)$ ($g=N+\mu+\nu$ denotes the total number of oscillators sent by Alice) and sends the state $\pi S \pi^*$ to Bob.

$\#2'$ Let the cp-map $T:\mathscr{B}(\mathscr{H}^{\otimes g}) \to \mathscr{B}(\mathscr{H}^{\otimes g})$ denote the quantum channel between Alice and Bob. T represents the (possibly collective) eavesdropping strategy used by Eve, $\mathscr{H}$ is the Hilbert space of an oscillator and $\mathscr{B}(\mathscr{H}^{\otimes g})$ the space of bounded operators on $\mathscr{H}^{\otimes g}$.

$\#3$ After Bob acknowledges receipt, Alice reveals $\pi$ and Bob undoes the permutation:
$T(\pi S \pi^*) \to T^{\pi}(S) \equiv \pi^*T(\pi S \pi^*)\pi$. Also, $\forall j=1 \ldots \mu$, Alice discloses the values of $x_j=\textrm{tr}(\hat{x}\; \gamma_{\textrm{c}}(j))$, and $\forall j=1 \ldots \nu$, Alice discloses the values of $p_j=\textrm{tr}(\hat{p}\; \sigma(j))$.

$\#4$ Bob measures the following effects:
\beqa
X^{(N+j)}(x_j) \equiv \gras{1}^{\otimes N+j-1} \otimes X(x_j) \otimes \gras{1}^{\otimes g-(N+j)}; \quad j=1 \ldots \mu;\\
P^{(N+\mu+j)}(p_j) \equiv \gras{1}^{\otimes N+\mu+j-1} \otimes P(p_j) \otimes \gras{1}^{\otimes g-(N+\mu+j)}; \quad j=1 \ldots \nu,
\eeqa
where $X(x_j)=\int_{\sqrt{\pi}/2\alpha+x_j}^{+\infty} dx \; \ket{x}\bra{x}+\int_{-\infty}^{-\sqrt{\pi}/2\alpha+x_j} dx \; \ket{x}\bra{x}$ and  $P(p_j)=\int_{\sqrt{\pi}\alpha/2+p_j}^{+\infty} dp \; \ket{p}\bra{p}+\int_{-\infty}^{-\sqrt{\pi}\alpha/2+p_j} dp \; \ket{p}\bra{p}$. 
N.B. each of these measurement has a yes/no outcome. Also note that Bob can as well measure these effects by performing a homodyne measurement on the corresponding oscillators, since after the measurements he no more needs these oscillators.

$\# 5$ Let $e_b(j)$ denote the outcome of the $j$th measurement, $j=1 \ldots \mu$ $[ e_b(j)=1$ if the effect $X^{(N+j)}(x_j) ]$ is measured and $e_b(j)=0$ otherwise). Likewise, we define $e_p(j)$, $j=1 \ldots \nu$. If the estimates for the bit error rate and phase error rate,  $e_b=\frac{1}{\mu}\sum_j e_b(j)$ and $e_p=\frac{1}{\nu}\sum_j e_p(j)$ respectively, satisfy the CSS rate inequality (\ref{eq:CSSrate}), Alice and Bob proceed, as described above, to distill a secret key from the remaining oscillators $S_{\textrm{key}}$.

\subsection{Estimation of the phase-error rate without squeezing}

To convert this last protocol to a secure coherent-state protocol, all we need is to prove that the phase error rate, $e_p$, can be estimated upon Alice sending only coherent states instead of squeezed states and Bob performing only homodyne measurements. In the following, $g$ will denote again the total number of oscillators sent by Alice and $M$ will denote a ("sufficiently large") integer.

Let $S_m$  and $S'_m$, $m=1 \ldots M$, denote two arrays of oscillators, each in a coherent state. Suppose that for all squeezed states $\sigma(j)$, $j=1 \ldots \nu$, involved in the last protocol, Alice and Bob had a means to estimate the quantities 
\beq\label{eq:sandwich}
\phi(j)=\frac{1}{M}\sum_{m=1}^{M} \textrm{tr}(T^{\pi}(S_m \otimes \sigma(j) \otimes S'_m)P^{(|S_m|+1)}(p_j)).
\eeq
Then, the quantity $\Phi=\frac{1}{\nu}\sum_{j=1}^{\nu}\phi(j)$ would certainly be as reliable an estimator for the phase error rate as the quantity $e_p$ Alice and Bob get in the modified squeezed state protocol. Now consider a situation where the first four steps of the last protocol are replaced by the following:

$\# 1$ Alice prepares the state 
\bed
R=S^{\textrm{coh}} \otimes S_{\text{ck}}^{\text{p}}\text{, with } S_{\text{ck}}^{\text{p}} = \bigotimes_{k=1}^{K} \gamma_k^{\otimes M},
\eed
where $S^{\textrm{coh}}$ is given by Eq.(\ref{eq:defscoh}) and where $\{\gamma_k\}$ denotes $K$ different coherent states, which will be used instead of the $p$-squeezed states to estimate $e_p$. 

$\# 2$ Alice picks a random permutation $\pi \in \textrm{Sym}(g)$ and sends the state $ \pi R \pi^*$ to Bob through the quantum channel. 

$\#2'$ Eve acts collectively: $\pi R \pi^* \to T(\pi R \pi^*)$. 

$\#3$ Bob acknowledges receipt of the oscillators, Alice reveals $\pi$ and Bob undoes the permutation $T(\pi R \pi^*) \to T^{\pi}(R)=\pi^* T(\pi R \pi^*) \pi$. Also, $\forall j=1 \ldots \mu$, Alice reveals the values $x_j=\textrm{tr}(\hat{x} \gamma_c(j))$, and $\forall j=1 \ldots \nu$, she reveals the values $p_j=\textrm{tr}(\hat{p} \sigma(j))$. 

$\# 4$ $\forall j=1 \ldots \mu$, Bob measures the effects $X^{(N+j)}(x_j)$, and $\forall k=1 \ldots K, m=1 \ldots M$, Bob measures the effects $P^{(N+\mu+(k-1)M+m)}(p_j)$. 

Note that since Bob performs individual measurements, the action of Eve's collective channel is the same as if she were acting with \emph{individual} channels $\tau^{\pi}_i, i=1 \ldots g$ ($g=N+\mu+KM$ denotes the total number of oscillators sent by Alice) defined, in Schr\"odinger picture, by

\begin{equation}
\tau^{\pi}_i:\mathscr{B}(\mathscr{H}) \to \mathscr{B}(\mathscr{H}):
\rho \to 
\textrm{Tr}_{'i'}T^{\pi}( \gamma^{(i-1)} \otimes \rho \otimes \gamma^{(g-i)}),
\end{equation}
for $i=1 \ldots g$, where $\textrm{Tr}_{'i'}$ denotes the partial trace over all subsystems but the $i$th, and 
$\gamma^{(i-1)}$ (resp. $\gamma^{(g-i)}$) represents an array of $(i-1)$ (resp. $(g-i)$) consecutive coherent states from the $g$ coherent states $S_{\text{key}} \otimes S_{\text{ck}}^{\text{b}} \otimes S_{\text{ck}}^{\text{p}}$.

%\bed
%\tau^{\pi}_i:\mathscr{B}(\mathscr{H}) \to \mathscr{B}(\mathscr{H}):
%\rho \to 
%\textrm{Tr}_{'i'}T^{\pi}(S_{\textrm{key}} \otimes \bigotimes_{h=1}^{i-N-1}\gamma_c(h) \otimes \rho \otimes \bigotimes_{h=i-N+1}^{\mu}\gamma_c(h) \otimes \bigotimes_{k=1}^{K} \gamma_k^{\otimes M}),
%\eed
%for $N+1 \leq i \leq N+\mu$, and  
%\bed
%\tau^{\pi}_i:\mathscr{B}(\mathscr{H}) \to \mathscr{B}(\mathscr{H}):
%\rho \to 
%\textrm{Tr}_{'i'}T^{\pi}(S^{\textrm{coh}} \otimes \bigotimes_{k' < k} \gamma_{k'}^{\otimes M} \otimes \gamma_k^{\otimes m-1} \otimes \rho \otimes \gamma_k^{\otimes M-m} \otimes \bigotimes_{k' > k} \gamma_{k'}^{\otimes M}),
%\eed
%for $i(k,m)=N+\mu+(k-1)M+m$, with $1 \leq k \leq K$ and $1 \leq m \leq M$.

Thus, in step $\# 4$, everything happens as if Bob were measuring $P(p_j)$ on $\tau_i^{\pi}(\gamma_k)$, $i(k,m)=N+\mu+(k-1)M+m$. Let $f(i(k,m),j,k)$ denote the outcomes he obtains  and let $F(j,k)$ denote the mean value of these $M$ outcomes. Due to the random permutation $\pi$, if $M$ is sufficiently large, we can be statistically confident that $F(j,k)$ doesn't depend on the individual cp-map indices, $i$, chosen for the test, i.e. we can be statistically confident that for all $M$-uple of indices $1 \leq i_1 \ldots i_M \leq N+\mu+KM$, would have Bob measured $P(p_j)$ on $\tau_{i_1}^{\pi}(\gamma_k) \ldots \tau_{i_M}^{\pi}(\gamma_k)$, the obtained outcomes $f(i_1,j,k) \ldots f(i_M,j,k)$ would have been such that $\frac{1}{M}\sum_{m=1}^{M}f(i_m,j,k) \approx F(j,k)$.

In particular, we can be statistically confident that
\bed
\frac{1}{M}\sum_{m=1}^{M}f(i(1,m),j,k) \approx F(j,k),
\eed
and a fortiori that
\bed
\frac{1}{M} \sum_{m=1}^{M} \textrm{tr} [P(p_j) \tau^{\pi}_{N+\mu+m}(\gamma_k)] \approx F(j,k).
\eed
That is,
\beq
\frac{1}{M}\sum_{m=1}^{M}\textrm{tr} [P^{(N+\mu+m)}(p_j) T^{\pi}(S^{\textrm{coh}} \otimes \gamma_1^ {\otimes m-1} \otimes \gamma_k \otimes \gamma_1^{M-m} \otimes \bigotimes_{k' > 1} \gamma_{k'}^{\otimes M})] \approx F(j,k).
\eeq

Defining $S_m \equiv S^{\textrm{coh}} \otimes \gamma_1^{\otimes m-1}, \;
S'_m \equiv \gamma_1^{\otimes M-m} \otimes \bigotimes_{k' > 1} \gamma_{k'}^{\otimes M}$, the latter equation reads 
\beq\label{eq:itconverges}
\frac{1}{M}\sum_{m=1}^M \tr (P^{(N+\mu+m)}(p_j) T^{\pi}(S_m \otimes \gamma_k \otimes S'_m)) \approx F(j,k).
\eeq
Now let us introduce the operator 
\bed
\mathscr{E}_j=\frac{1}{M}\sum_m \Tr_{\mathscr{H}_m} \Tr_{\mathscr{H'}_m} (T_*^{\pi}(P^{(N+\mu+m)}(p_j))(S_m \otimes \gras{1} \otimes S'_m)),
\eed
where $T_*^{\pi}$, the dual of $T^{\pi}$ defines evolution in Heisenberg picture \cite{keyl02} (it is related to $T^{\pi}$ through the identity $\tr(T^{\pi}(\rho)A)=\tr(\rho T^{\pi}_*(A))$), and where $\mathscr{H}_m$ (resp. $\mathscr{H'}_m$) denotes the Hilbert space supporting the state $S_m$ (resp. $S'_m$). With the help of $\mathscr{E}_j$, Eq.(\ref{eq:itconverges}) can be re-written as $\tr \mathscr{E}_j \gamma_k=F(j,k)$ and similarly Eq.(\ref{eq:sandwich}) as $\phi(j)=\tr \mathscr{E}_j \sigma(j)$. It now only remains to prove that $\tr \mathscr{E}_j \sigma(j)$ can be inferred from the quantities $F(j,k)$ when the coherent states $\gamma_k$ are correctly chosen.

Let $\sigma(j)=\ket{\psi^j}\bra{\psi^j}$ and let $\sum_n \psi_{n}^j \ket{n}$ denote the expansion of $\ket{\psi^j}$ in Fock basis. Since $\sum_n |\psi_{n}^j|^2=1$, we have $\forall \epsilon > 0, \exists \mathscr{N}_j$ s.t. $\sum_{n=\mathscr{N}_j+1}^{\infty} |\psi^j_n|^2 < \epsilon$. Let $\mathscr{N}=\textrm{max}_j \mathscr{N}_j$ and let us denote $\ket{\psi^j_\mathscr{N}}=\sum_{n=0}^\mathscr{N} \psi^j_{n} \ket{n}$ and $\ket{\psi_\mathscr{N}^{j,c}}=\ket{\psi^j}-\ket{\psi^j_\mathscr{N}}$. We have
\bed
|\bra{\psi^j} \mathscr{E}_j \ket{\psi^j}-\bra{\psi^j_{\mathscr{N}}} \mathscr{E}_j \ket{\psi^j_{\mathscr{N}}}| < \epsilon+2\sqrt{\epsilon}.
\eed
Indeed, $0 \leq \mathscr{E}_j \leq \gras{1}$ and Cauchy-Schwarz inequality imply that $\bra{\psi^{j,c}_{\mathscr{N}}} \mathscr{E}_j \ket{\psi^{j,c}_{\mathscr{N}}} \leq || \psi^{j,c}_{\mathscr{N}} ||^2 < \epsilon$ and that $|\bra{\psi^j_{\mathscr{N}}} \mathscr{E}_j \ket{\psi^{j,c}_{\mathscr{N}}}| \leq  \sqrt{\epsilon}$. Thus the knowledge of $\bra{\psi^j_{\mathscr{N}}} \mathscr{E}_j \ket{\psi^j_{\mathscr{N}}}$ brings (in arbitrarily good approximation) the knowledge of $\bra{\psi^j} \mathscr{E}_j \ket{\psi^j}$. Also, the quantities $\bra{\psi^j_{\mathscr{N}}} \mathscr{E}_j \ket{\psi^j_{\mathscr{N}}}$ can be inferred from the $(\mathscr{N}+1)^2$ quantities $\bra{l} \mathscr{E}_j \ket{n}, 0 \leq l,n \leq \mathscr{N}$. It thus only  remain to show how to estimate these quantities. Let $\ket{\alpha_k}$ denote $(\mathscr{N}+1)^2=K$ coherent states and let $\sum_{n=0}^{\infty} c_k^n \ket{n}$ denote their expansions in Fock basis. The states $\ket{\alpha_k}$ are chosen such that 
$\sum_{n=N+1}^{\infty} |c_k^n|^2 < \epsilon, \forall k=1 \ldots K$. Thus setting $\gamma_k=\ket{\alpha_k}\bra{\alpha_k}$, we have 
\beq\label{eq:almostend}
F(j,k)+\eta_k=\sum_{l,n=0}^{N} c_k^{l*} c_k^n \bra{l}\mathscr{E}_j\ket{n},
\eeq
where $|\eta_k| < \epsilon+2\sqrt{\epsilon}$. It is always possible to choose the  values $\alpha_k$ such that the matrix $\Gamma=[\Gamma_k^{(l,n)}] \equiv c_k^{l*} c_k^{n}$ is invertible. Indeed, $\textrm{det} \Gamma$ is of the form $e^{-\frac{1}{2} \sum_{k=1}^{K}|\alpha_k|^2} Q(\alpha_1,\ldots,\alpha_K;\alpha_1^*,\ldots,\alpha_K^*)$, where $Q$ is of the form $\sum_{\lambda,\mu=0}^{N} Q^1_{\lambda,\mu}(\alpha_2,\ldots,\alpha_K;\alpha_2^*,\ldots,\alpha_K^*) \alpha_1^{\lambda} \alpha_1^{*\mu}$. Unless all coefficients $Q^1_{\lambda,\mu}$ are identically zero, $\textrm{det} \Gamma$ only vanishes for a finite number of values of $\alpha_1$ when all the values $\alpha_2,\ldots,\alpha_K$ are fixed. Similarly, each polynomial $Q^1_{\lambda,\mu}$ is of the form $\sum_{\nu,\omega} Q^2_{\nu,\omega}(\alpha_3,\ldots,\alpha_K;\alpha_3^*,\ldots,\alpha_K^*) \alpha_2^{\nu} \alpha_2^{*\omega}$, so that (again) unless all the coefficients $Q^2_{\nu,\omega}$ are identically zero, $Q^1_{\lambda,\mu}$ only vanishes for a finite number of values of $\alpha_2$ when all the values $\alpha_3,\ldots,\alpha_K$ are fixed. And so on, continuing the reasoning, we see that we can always meet the condition $\textrm{det} \Gamma \neq 0$ upon fixing $\alpha_K$ to some value such that not all $Q_{\nu,\omega}^{K-1}(\alpha_K;\alpha_K^*)$ are zero, fixing $\alpha_{K-1}$ to some value such that not all $Q_{\nu,\omega}^{K-2}(\alpha_{K-1},\alpha_K;\alpha_{K-1}^*,\alpha_K^*)$ are zero, $\ldots$, and fixing $\alpha_1$ such that $Q(\alpha_1,\ldots,\alpha_K;\alpha_1^*,\ldots,\alpha_K^*) \neq 0$. We see that inverting the relation Eq.(\ref{eq:almostend}) will provide estimates for the quantities $\bra{l}\mathscr{E}_j\ket{n}$ (from which we infer the quantities $\tr\mathscr{E}_j \sigma(j)$), and hence the phase error rate $\Phi$. It is however important to note that these estimates will only be accurate if $||\Gamma^{-1}\gras{\eta}|| \ll ||\Gamma^{-1} \gras{F}(j)||$, where $\gras{\eta}=(\eta_0,\ldots,\eta_K)$ and $\gras{F}(j)=(F(j,0),\ldots,F(j,K))$. Therefore, Alice and Bob should add an additional step in the protocol where they check that this latter condition is indeed satisfied. %A crude way to do that would be to test whether $||\Gamma^{-1}|| K (\epsilon+2\sqrt(\epsilon)) << ||\Gamma^{-1} \gras{F}(j)||$. %This point will be discussed in more details in \cite{vana03:qslices}

%To estimate the latter quantities, we introduce the pseudo-mixtures of coherent states 
%\bed
%\Gamma^n(r)=\int \frac{d\theta}{2\pi} e^{in\theta} \ket{r e^{i \theta}} \bra{r e^{i\theta}}=e^{-r^2} \sum_{l=0}^{\infty} \frac{r^{2l+n}}{\sqrt{l! (l+n)!}} \ket{l}\bra{l+n}, \; r \in \gras{R}^+.
%\eed
%The quantities $\tr \Gamma^n(r) \mathscr{E}_j$ can be estimated from the quantities $F(j,k)$ which are estimated with arbitrarily high statistical confidence upon Alice sending coherent states and Bob performing $\hat{p}$ homodyne measurements. Thus, considering various values of $r$, the quantities $\bra{l} \mathscr{E}_j \ket{l+n}$ can be inferred. 

\section{Further extensions}

Let us now sketch how this protocol can extended by relaxing the translational symmetry of the qubit encoding scheme so as to improve its efficiency and get closer to the effective key extraction procedure introduced in \cite{vana01} and implemented in \cite{gros03}. The technical details will be described in a forthcoming paper \cite{vana03:qslices}. In the same way we can split a real number into an integer and a fractional part as in Eq.(\ref{eq:xdecomposed}), we can encode $m$ qubits in an oscillator with the following linear transformation (up to normalization)
\begin{equation}
\ket{x} \to \ket{\bar{S}(x)}_{\sysid{\bar{s}}} \otimes \ket{S_1(x)}_{\sysid{s}_1} \otimes \ldots \otimes \ket{S_{m}(x)}_{\sysid{s}_m},
\end{equation}
which decomposes the oscillator into $m$ qubit subsystems $\sysid{s}_i$, $i=1 \dots m$, and some continuous subsystem $\sysid{\bar{s}}$. The functions $S_i(x)$, $i=1 \dots m$, called slices, have range $\{0,1\}$. They can be general, and in particular the bit values 0 and 1 do not need be placed periodically on the $x$ axis as in \cite{gott01:crypto}. Instead, one can imagine the $x$ axis cut into $2^m$ intervals, each assigned to $m$ binary values $S_{1 \dots m}(x)$ (see \cite{vana01} for more details). Then, $\bar{S}(x)$ (with range $[0;1]$) is a continuous function that carries the remaining continuous information about $x$ not contained in $S_{1\dots m}(x)$. Note the similarity with Eq.(\ref{eq:xdecomposed}), where the bit value is in the parity of $S(x)$ and the remaining continuous information in $\bar{S}(x)$. Similarly, Bob can also decompose his oscillator into $m$ qubits ($\sysid{e}_i$, $i=1 \dots m$) and one continuous subsystem ($\sysid{\bar{e}}$). These decompositions are chosen so as to get high entanglement purification rates $R_i = 1-h(e^b_i)-h(e^p_i)$ for the pairs $\sysid{s}_i\sysid{e}_i$, $i=1 \dots m$. The rate $R_i$ of each slice is calculated by tracing out the other parts of the system, that is, as if the other slices are controlled by Eve. Consequently, the total number of secret bits produced per oscillator can be safely obtained by summing over the slices, $R=\sum_{i=1}^{m} R_i$. Following an argument similar to the one above, the bit and phase error rates, $e^b_i$ and $e^p_i$, can be estimated from homodyne detection with arbitrary precision.

To further improve the efficiency, the decomposition into the subsystems $\sysid{e}_{1\dots m}$ may depend both on $\bar{S}(x)$ (e.g., like in the above protocol, where Bob uses $\bar{S}(x)$ to re-adjust his measured value) and on the bit value of the previous slices $S_{j<i}(x)$. %This is possible since the bit syndromes $\xi^b_j$ sent by Alice, to perform CSS error correction, can be given to Bob prior to doing his mapping. In the derived QKD protocol,
This allows Bob to estimate Alice's bits $S_i(x)$ with the information already acquired from the correction of the bits $S_{j<i}(x)$, thereby improving the correlations \cite{vana03:qslices}.

For illustration, we have applied this ``sliced" encoding to the special case of the noiseless attenuation channel, which is of practical importance. This corresponds to the simplest attack where Eve puts a beam-splitter in between two sections of a lossless line, sending vacuum at the second input port. Similarly to the implementation of \cite{gros03}, the modulation variance of Alice was chosen equal to $31\times~\text{the vacuum noise}$ in both quadratures, which gives Alice and Bob up to $2.5$ common bits in the absence of losses. Using this encoding with two slices, we were able to get the net key rates described in Table~\ref{fig:tac}. The slices $S_1$ and $S_2$ are defined by dividing the real axis into four equiprobable intervals labeled by two bits, with $S_1$ ($S_2$) being the least (most) significant bit. For the case with no losses, it is thus possible to distill $R=0.752+0.938=1.69$ secret bits per oscillator, thus significantly improving the rate obtainable with the encoding of \cite{gott01:crypto}, namely, up to one bit per oscillator. Due to the higher bit error rate, it was not possible to distill secret bits in slice~1 with losses beyond 0.7~dB. It was however still possible to distill secret bits in slice~2 up to 1.4~dB losses (about 10~km with fiber optics with losses of 0.15~dB/km). This value for the maximum tolerable loss does not stem from any fundamental reason, and might be improved by tuning the modulation variance and/or the functions $S_1$ and $S_2$.

\begin{table}[t]
\begin{center}
\begin{tabular}{|r|r|r|r|r|r|r|}
\hline
& \multicolumn{3}{|c|}{$\sysid{s}_1\sysid{e}_1$} & \multicolumn{3}{|c|}{$\sysid{s}_2\sysid{e}_2$} \\
\cline{2-7}
Losses & $e^b_1$ & $e^p_1$ & $R_1$ & $e^b_2$ & $e^p_2$ & $R_2$ \\
\hline
0.0 dB & 3.11\% & 5.33\% & 0.752 & 0.0000401 & 0.710\% & 0.938 \\
0.4 dB & 3.77\% & 13.7\% & 0.193 & 0.0000782 & 28.6\% & 0.135 \\
0.7 dB & 4.32\% & 20.0\% & 0.0204 & 0.000125 & 37.5\% & 0.0434 \\
\cline{2-4}
1.0 dB & \multicolumn{3}{|c|}{-} & 0.000194 & 42.3\% & 0.0147 \\
1.4 dB & \multicolumn{3}{|c|}{-} & 0.000335 & 45.6\% & 0.00114 \\
\hline
\end{tabular}
\end{center}
\caption{Bit and phase error rates and corresponding net key rates for a two-slice encoding as a function of the channel attenuation.}
\label{fig:tac}
\end{table}

\section{Conclusions}

In summary, we have studied the security of Gaussian-modulated coherent-state protocols against arbitrary attacks. We have shown how to extend the protocol of \cite{gott01:crypto} to remove the need of squeezing for estimating the phase error rate. This quantity can also be estimated using coherent states modulated in two conjugate quadratures, homodyne measurements, and appropriate classical post-processing. The equivalence between the derived coherent-state QKD protocol and a squeezed state protocol, itself equivalent to a protocol based on EPR purification with CSS codes, assesses the security against arbitrary attacks, including collective and/or non-Gaussian attacks.
This is compatible with the fact that coherent-state protocols have been shown to be equivalent to protocols involving entangled bipartite Gaussian states \cite{gros03:virtual}. Also, we extended the encoding scheme of \cite{gott01:crypto} in order to improve the net key rate. By transmitting more than one bit per oscillator in a manner derived from an effective reconciliation procedure \cite{vana01,gros03}, numerical results have shown that a secret key can be extracted in an attenuation channel with significant loss. 
%This study naturally leads to some open questions. Our security analysis relies on the assumption that all errors, such as those due to finite squeezing , are attributed to an eavesdropper. 

%For example, can we use the same techniques to derive a secure protocol based on reverse reconciliation \cite{gros03}? This may allow a better robustness to perturbations.

\section{Acknowledgments}

We thank P. Grangier, F. Grosshans, P. Navez for discussions, and especially J. Preskill for fruitful correspondence. S.I. acknowledges support from the Belgian FRIA foundation. G.V.A. and N.J.C. acknowledge financial support from the Communaut\'e Fran\c caise de Belgique under grant ARC 00/05-251, from the IUAP programme of the Belgian government under grant V-18 and from the EU under project RESQ (IST-2001-35759).

\bibliography{qit}

\end{document}